\documentclass[twocolumn,10pt,a4paper]{article}
\usepackage[top=2cm,right=2cm,left=2cm,bottom=2cm]{geometry}
\usepackage{helvet,times}
\usepackage{bm,textcomp}
\usepackage[affil-it]{authblk}
\usepackage{graphicx}
\usepackage{amsmath}
\usepackage{amsthm}
\usepackage{amsfonts}
\usepackage{pstricks}
\usepackage{tikz}
\usetikzlibrary{arrows,decorations.pathmorphing,backgrounds,positioning,fit,petri,arrows.meta}
\usepackage{pgfplots}
\pgfplotsset{compat=1.14}
\usepackage{cuted}

\newcommand{\corresponding}{\hat{=}}
\newcommand{\plaind}{\mathrm{d}}

\title{Is Actin Filament and Microtubule Growth Reaction- or Diffusion-Limited?}
\author{Johannes Pausch\thanks{}, Gunnar Pruessner}

\affil{Department of Mathematics, Imperial College London, UK}
\begin{document}
\setcounter{page}{1}
\twocolumn[
  \begin{@twocolumnfalse}
\maketitle
\begin{abstract}%
{Inside cells of living organisms, actin filaments and microtubules self-assemble and dissemble dynamically by incorporating actin or tubulin from the cell plasma or releasing it into their tips' surroundings. Such reaction-diffusion systems can show diffusion- or reaction-limited behaviour. However, neither limit explains the experimental data: while the offset of the linear relation between growth speed and bulk tubulin density contradicts the diffusion limit, the surprisingly large variance of the growth speed rejects a pure reaction limit. In this Letter, we accommodate both limits and use a Doi-Peliti field-theory model to estimate how diffusive transport is perturbing the chemical reactions at the filament tip. Furthermore, a crossover bulk density is predicted at which the limiting process changes from chemical reactions to diffusive transport. In addition, we explain and estimate larger variances of the growth speed. }
\end{abstract}
\end{@twocolumnfalse}
]
{
\renewcommand{\thefootnote}%
    {\fnsymbol{footnote}}
  \footnotetext[1]{corresponding author: j.pausch15@imperial.ac.uk}
  }
Microtubules and actin filaments are structures that polymerise by incorporating and releasing their diffusively moving building blocks, tubulin and actin. They are responsible for growth, shape, movement, and transport processes, among others, and can span the entire cell \cite{Boal2012}.  Their dynamics have been studied intensively experimentally\footnote{\label{Footnote:Experiments}See Table S1 in the Supplemental Information of \cite{Gardner2011} for microtubule assembly rates. For actin filaments, assembly rates can be found in \cite{Pollard1984,Fujiwara2002,Pollard2005,Pollard2007}.} and theoretically.\footnote{Theoretical work includes \cite{Mitchison1984,Frieden1985,Carlier1987,Chen1987,Walker1988,Bayley1993,Odde1997} for microtubules and \cite{Frieden1985,Fujiwara2002,Guo2009} for actin filaments.} However, many questions about their dynamics remain debated or completely unanswered. This includes: Is their growth limited by diffusion to their  tip \cite{Odde1997,Drenckhahn1986} or by the chemical reaction rates for incorporation \cite{Bayley1993,Drenckhahn1986}? How can the large variance of their growth be explained \cite{Gardner2011}?

In experiments, the filament growth speed $\langle v\rangle$ can be measured as a function of the bulk density $\zeta$ of tubulin or actin. An effective incorporation coefficient $k_\text{on}$ and effective release rate $k_\text{off}$ are determined as parameters of a linear fit of growth speed data over bulk concentration $\zeta$\begin{align} \label{Eq-Experiment-Speed}
\langle v\rangle=h(k_\text{on}\zeta-k_\text{off}),
\end{align}
where $h$ is the effective growth length per incorporated particle. There are two mean field approaches for modeling the filament growth speed $\langle v\rangle$. 

The first approach assumes that diffusive transport is quicker than the chemical reactions \cite{Mitchison1984,Carlier1987,Chen1987,Walker1988,Bayley1993} and that polymerisation is therefore \textit{reaction-limited}. This effectively infinite diffusivity $D$ implies that particle concentration is homogeneous, and in particular, that there is no significant depletion close to the filament tip. Thus, filament growth is determined by two Poisson processes, incorporation with rate $\lambda\zeta$ and release with rate $\tau$, each of which is associated with a step of length $h$. The two competing processes create a Skellam distribution \cite{Skellam1946} with expected growth speed $\langle v\rangle_R$ and effective diffusion constant $D_\text{eff}$ \begin{align}\label{Eq-R-Limit}
\langle v\rangle_R=h(\lambda\zeta-\tau)\quad\text{and}\quad D_\text{eff}=h^2(\lambda\zeta+\tau).
\end{align} 
However, in comparison with experiments \cite{Gardner2011,Fujiwara2002,Pollard2005}, $D_\text{eff}$ is too small. Furthermore, it implies an independence of the growth speed from the viscosity of the medium, which was rejected experimentally for actin filaments \cite{Drenckhahn1986} and microtubules \cite{Wieczorek2013}. Thus, a purely reaction-limited behaviour is rejected.

The second approach assumes that transport by diffusion is slower than the chemical reactions \cite{Drenckhahn1986,Odde1997} and that self-assembly is therefore \textit{diffusion-limited}. This implies that, due to the incorporation into the tip, the building blocks are depleted locally.  The growth speed is determined by the diffusive flux to the tip. If the protein concentration $c(x,t)$ follows a steady state diffusion equation $0=D\Delta c(x,t)$, the particle flux $J$ through the absorbent reaction sphere of radius $R$ and the growth length $h$ determine the growth speed $\langle v\rangle_D$, which is equivalent to Smoluchowski coagulation \cite{Smoluchowski1917}\begin{align}\label{Eq-Diff-Limit}
\langle v\rangle_D=Jh=4\pi DRh\zeta.
\end{align}
However, this limit cannot accommodate a release rate as any released particle would immediately be reabsorbed before diffusion can transport it away from the reaction surface. Furthermore, using this approach with typical parameters of microtubule assembly, only a small reduction of the bulk tubulin density to $89\%$ is found at the reaction surface \cite{Odde1997}. It therefore is not completely absorbent, as is theoretically suggested in \cite{Collins1950}.  According to the Stokes-Einstein equation \cite{Sutherland1905} for small Reynolds numbers, diffusion-limited growth implies that viscosity and incorporation rate are inversely proportional \cite{Berg1985} without offset. Tested in \cite{Drenckhahn1986} and \cite{Wieczorek2013}, small offsets for the growth of microtubules and the barbed actin filament ends are found, while a significant offset is found for the pointed ends of actin filaments. Thus, filament growth cannot be perfectly diffusion-limited either. Furthermore, this model is not probabilistic and therefore, it is not clear how to derive a variance of the growth speed.

\begin{figure}
\begin{center}
\includegraphics[width=\columnwidth]{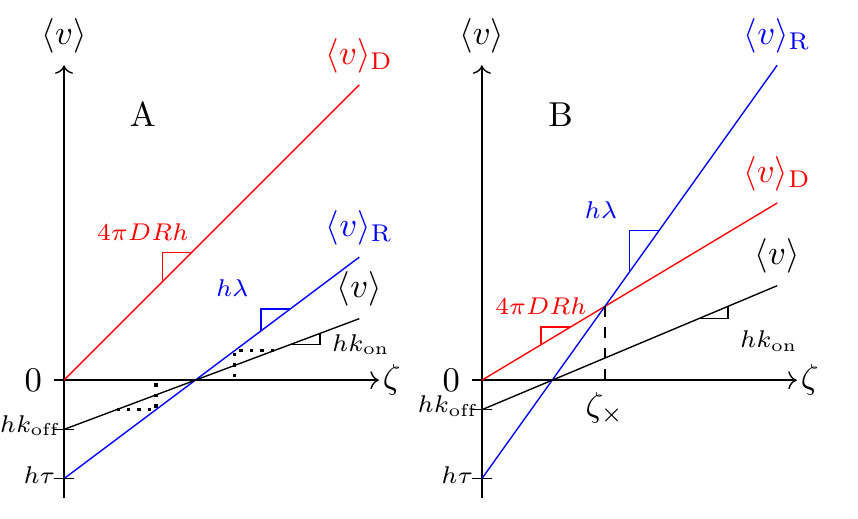}
\end{center}
\caption{Filament growth speed $\langle v\rangle$ as function of the bulk density $\zeta$ of its building blocks. The diffusion-limited $\langle v\rangle_D$ (red), the reaction-limited $\langle v\rangle_R$ (blue), and the real growth speed $\langle v\rangle$ (black) are shown. A: The system is reaction-limited, B: Reaction-limited growth changes to diffusion-limited growth at the crossover bulk density $\zeta_\times$. Local densities at the tip can be read off as the bulk density at which reaction-limited growth leads to the same speed (dotted line).}\label{Fig-Reaction-vs-diffusion-scheme}
\end{figure}

The diffusion- (Eq.~\eqref{Eq-Diff-Limit}) and reaction-limited (Eq.~\eqref{Eq-R-Limit}), as well as the measured growth speed are schematically depicted in Fig.~\ref{Fig-Reaction-vs-diffusion-scheme}. Measured growth and shrinking speeds will always be slower than both limits. When shrinking, the chemically possible shrinking speed is not attained as diffusion slows down transport away from the tip, leading to a locally higher particle density. It can be read off the plot as the density at which the reaction-limited is equal to the measured speed (dotted line). Analogously, when growing, the reaction-limited speed is not attained either because diffusion fails to maintain the bulk particle density around the tip.  The local density at the tip can be read off again as the density at which the reaction-limited speed is equal to the measured one (dotted line). In principle, there are three cases: the purely reaction limited case A, the mixed case B, where a crossover from reaction-limited to diffusion-limited behaviour occurs at the crossover bulk density $\zeta_\times$, and case C (not shown), where shrinking does not occur and the growth is limited by diffusion at all particle densities.   

Theoretically, progress can be made by going beyond mean field theory which allows to calculate how chemical reactions are perturbed by diffusive transport of the reactants. Here, filament self-assembly  is modelled on a three-dimensional lattice and described by a master equation (Supplementary Material, Eq.~A2). Following work by Doi \cite{Doi1976} and Peliti \cite{Peliti1985}, this model is transformed into a field theory. The derivation of the field theory is outlined briefly in the Supplementary Material, in Sections A2 and~A3.

\begin{figure}
\begin{center}
\includegraphics[width=\columnwidth]{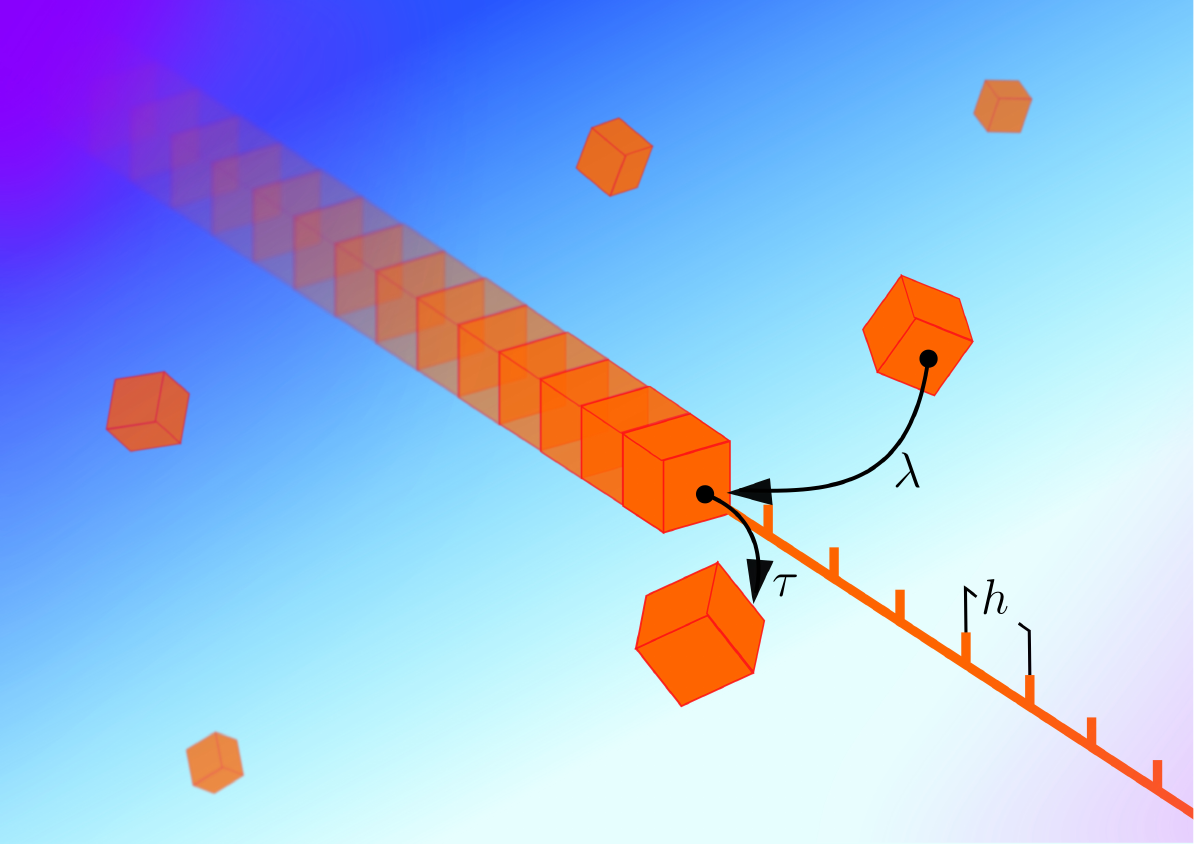}
\end{center}\caption{Schematic of the filament self-assembly process. Its building blocks move diffusively in $\mathbb{R}^3$. The filament tip is on a lattice with spacing $h$. Particles can be incorporated in the filament (coefficient $\lambda$) and released by it (rate $\tau$).}
\label{Fig-Schematic}
\end{figure}

In our model, the building blocks and the filament tip are represented by fields, which are interpreted as time-dependent probability distributions of their positions. Due to the field representation, they do not have a size. However, the finite size of the proteins is an important element of the step-wise filament growth. Therefore, our model is set up on a hybrid, three dimensional space: While particles move in continuous space $\mathbb{R}^3$, the filament tip is restricted to a discrete line $\mathbb{Z}$ with lattice constant $h$, overlaying the $z$-axis in $\mathbb{R}^3$, see Fig.~\ref{Fig-Schematic}. The persistence length $P$ and flexural rigidity $K$ of the filament we model are thus effectively infinite. This is a good approximation for microtubules ($P\approx5200\mu m$, $K\approx 2\cdot10^{-23}Nm^2$), while for actin filaments, this approximation is slightly worse ($P\approx18\mu m$, $K\approx7\cdot10^{-26}Nm^2$) \cite{Gittes1993}.  As we are modelling only a single polymer instead of the 13 microtubule protofilaments or the 2 actin filament strands, we interpret the lattice spacing $h$ as the effective growth step length. 

Each of the fields exists as an annihilation field and a creation field: $\varphi(x,t)$ and $\varphi^\dagger(x,t)$ for the building blocks, as well as  $\psi_j(t)$ and $\psi_j^\dagger(t)$ for the filament tip.  The creation field initiates a single particle or filament tip at the specified position and time, whereas the annihilation field measures their number at the point stated. Creation fields often appear as Doi-shifted fields \cite{Doi1976}, e.g. $\varphi^\dagger(x,t)=\widetilde\varphi(x,t)+1$. In addition, the particle annihilation field is shifted to measure deviations from the bulk density $\zeta$, i.e. $\varphi(x,t)=\check\varphi(x,t)+\zeta$.  Between creation and annihilation, the system evolves by the stochastic processes included in the model.  

There are six microscopic processes in our model. The units of the corresponding coefficients, denoted by $[\dots]$, are written as monomials of $T$ (time) and $L$ (length).\begin{itemize}\setlength\itemsep{0.2em}
\item particle diffusion with constant $D$, $[D]=T^{-1}L^{2}$
\item actin or tubulin absorption by the filament tip with coefficient $\lambda$ and subsequent movement of the tip by distance $h$ in the $+z$ direction, $[\lambda]=T^{-1}L^{3}$
\item particle release from the filament tip with rate $\tau$ and subsequent movement of the tip by distance $h$ in the $-z$ direction, $[\tau]=T^{-1}$
\item actin/tubulin creation with coefficient $\gamma$, $[\gamma]=T^{-1}L^{-3}$
\item extinction of actin/tubulin with rate $r$, $[r]=T^{-1}$
\item extinction of the filament tip with rate $\epsilon$, $[\epsilon]=T^{-1}$
\end{itemize}
Creation and extinction of the building blocks is balanced such that a constant bulk density $\zeta=\gamma/r$ is created. The two extinction processes are included in the field theory to enforce causality. After calculations, we let parameters $\gamma,r$ and $\epsilon$ tend to zero while keeping the ratio $\gamma/r=\zeta$ constant. Thus, the spontaneous extinction and creation are removed while a bulk density remains included.

All of the processes above are reflected in the action functional $\mathcal{A}$ of our model which splits up into a bilinear part and an interaction part $\mathcal{A}=\mathcal{A}_\text{lin}+\mathcal{A}_\text{int}$. The diffusion and extinction of particles is represented in the particle bilinear part\begin{align}\label{Eq-particle-prop}
\mathcal{A}_{\text{lin-P}}=\int\limits_{\mathbb{R}^4}\widetilde\varphi(x,t)\bigl(-\partial_t+D\Delta-r\bigr)\check\varphi(x,t)\plaind^3x\plaind t,
\end{align}
where $\Delta$ is the spatial Laplace operator.

The filament tip is stationary without the processes of incorporation or release of tubulin. It is described by\begin{align}\label{Eq-Fil-Prop1}
\mathcal{A}_{\text{lin-F-stat}}=\int\limits_{\mathbb{R}}\sum\limits_{j\in\mathbb{Z}}\widetilde\psi_j(t)\bigl(-\partial_t-\epsilon\bigr)\psi_j(t) \plaind t.
\end{align}

However, due to incorporation and release, the bilinear part includes jumps in steps of $1_z$\begin{align}\label{Eq-Fil-Prop2}\begin{split}
\mathcal{A}_{\text{lin-F-mov}}=\int\limits_{\mathbb{R}}\sum\limits_{j\in\mathbb{Z}}\Bigl(&\overbrace{\lambda\zeta\bigl(\widetilde\psi_{j+1_z}-\widetilde\psi_j\bigr)\psi_j}^{\text{growing}}\\&
+\underbrace{\tau\bigl(\widetilde\psi_{j-1_z}-\widetilde\psi_j\bigr)\psi_j \Bigr)}_{\text{shrinking}}\plaind t,\end{split}
\end{align}
where we omitted the time dependence of the fields for better readability. The first part  corresponds to filament growth, while the second part describes shrinking of the filament. Jumps on the lattice are indicated by $\pm1_z$.

All three bilinear actions together make up $\mathcal{A}_\text{lin}=\mathcal{A}_{\text{lin-P}}+\mathcal{A}_{\text{lin-F-stat}}+\mathcal{A}_{\text{lin-F-mov}}$.

$ $\begin{strip}The interaction part of $\mathcal{A}$ has the form\begin{align}\begin{split}
\mathcal{A}_\text{int}=&\int\limits_{\mathbb{R}}\sum\limits_{j\in\mathbb{Z}}\biggl[\lambda\Bigl(\underbrace{\bigl(\widetilde\psi_{j+1_z}-\widetilde\psi_j\bigr)\psi_j\check\varphi(hj)}_{(a)}
-\underbrace{\widetilde\psi_j\psi_j\widetilde\varphi(hj)\check\varphi(hj)}_{(b)}
-\underbrace{\psi_j\widetilde\varphi(hj)\check\varphi(hj)}_{(c)}\Bigr)\\
&\hspace{2.5cm}
+\underbrace{\bigl(\tau\widetilde\psi_{j-1_z}-\lambda\zeta\widetilde\psi_j\bigr)\psi_j\widetilde\varphi(hj)}_{(d)}
+\underbrace{(\tau-\lambda\zeta)\widetilde\varphi(hj)\psi_j}_{(e)}\biggr]\plaind t,
\end{split}\label{Eq-Interaction-Action}\end{align}\end{strip}
where the time dependency is omitted for better readability.

The different parts of the interaction describe the following processes: \begin{itemize}\setlength\itemsep{0.2em}
\item[(a)] the filament grows, i.e. the filament tip moves one step in the positive $z$ direction upon incorporating an actin/tubulin particle; 
\item[(b)] as particles are incorporated into the tip, its density is reduced locally, resulting in anticorrelations of the tip and the particle density;
\item[(c)] particle density is reduced by incorporation into the tip;
\item[(d)] in the presence of a tip, the particle density is increased by spontaneous release ($\tau$) and decreased by incorporation ($\lambda\zeta$), leading to corresponding correlations of tip and particle densities; 
\item[(e)] $\tau$: particle density is increased because the filament releases a particle; $\lambda\zeta$: particle density is decreased because the filament incorporates a particle.
\end{itemize}  

\begin{figure}[b!]
\begin{center}
\includegraphics[width=\columnwidth]{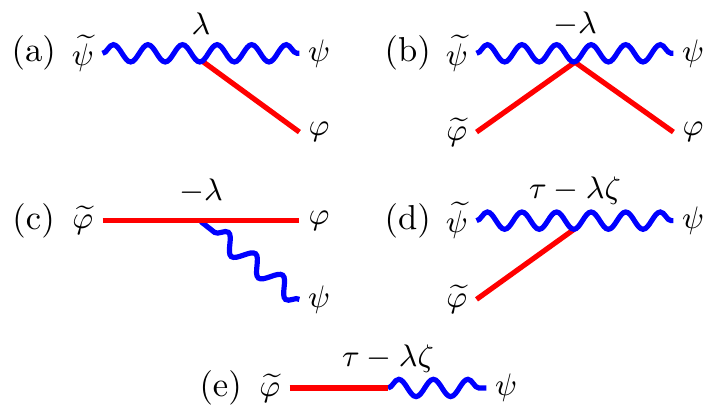}
\end{center}
\caption{The stochastic processes that appear in $\mathcal{A}_\text{int}$, Eq.~\eqref{Eq-Interaction-Action}, are represented as amputated vertices in Feynman diagrams. Curly blue lines represent filaments, while straight red lines stand for their building blocks, actin or tubulin. All Feynman diagrams are read from right to left.}
\label{Fig-Action}
\end{figure}

Field-theoretic propagations and interactions are schematically represented by Feynman diagrams. Particle propagation is drawn as a straight red line, filament propagation is depicted as a curly blue line, and interactions are illustrated as vertices, see Fig.~\ref{Fig-Action}.

The interplay between propagation and interaction in a system governed by the action $\mathcal{A}$ can be calculated using the path integral. The system may be initialised by placing a filament tip at position $hj_0=0$ at time $t_0=0$. Then, the system evolves and particle concentrations, the filament tip positions, or moments of their distributions can be measured at a later point in time. In general, if the observable that we want to measure is represented by a combination of fields $\mathcal{O}(t)$, then its time-dependent, spatial probability distribution is given by the following path integral (see e.g. \cite{Taeuber2014} for a detailed review)\begin{align}
\langle\mathcal{O}(t)\psi_0^\dagger(0)\rangle\hspace{-0.1cm}:=\hspace{-0.1cm}\sum\limits_{\ell=0}^{\infty}\hspace{-0.08cm}\int\hspace{-0.12cm}\mathcal{D}[\varphi,\psi]\mathcal{O}(t)\psi_0^\dagger(0)e^{-\mathcal{A}_\text{prop}}\frac{(-\mathcal{A}_{\text{int}})^\ell}{\ell!}.
\label{Eq-Path-Integral}
\end{align}
Formally, this integral is summing all variations of all fields involved of all stochastic processes possibly occurring. The path integral is normalised such that\begin{align}
\langle1\rangle=\sum\limits_{\ell=0}^{\infty}\hspace{-0.08cm}\int\hspace{-0.12cm}\mathcal{D}[\varphi,\psi]e^{-\mathcal{A}_\text{prop}}\frac{(-\mathcal{A}_{\text{int}})^\ell}{\ell!}=1.
\end{align}

The $\ell$-th term of the series is the contribution of all processes with $\ell$ interactions.\footnote{These interactions are in the field-theoretic sense. In fact, the term for $\ell=0$ already includes an arbitrary number of chemical interactions.} The distribution of the filament's tip position $j$ is given by $\langle\psi_j(t)\psi^\dagger_0(0)\rangle$. 
The expected growth speed and its variance are then determined by calculating the filament's expected position and its variance after time $t$.  In the  following, we consider three approximations of $\langle\psi_j(t)\psi^\dagger_0(0)\rangle$: Firstly, $\langle\psi_j(t)\psi^\dagger_0(0)\rangle_0$ is the reaction-limited distribution, cutting the sum in Eq.~\eqref{Eq-Path-Integral} at $\ell=0$, which results in only one Feynman diagram shown in Eq.~\eqref{Eq-Dyson-01}. Secondly, $\langle\psi_j(t)\psi^\dagger_0(0)\rangle_{2}$ is terminating the sum at $\ell=2$, resulting in two diagrams shown in Eq.~\eqref{Eq-Dyson-01} and~\eqref{Eq-Dyson-02}.  Thirdly, $\langle\psi_j(t)\psi^\dagger_0(0)\rangle_\text{Dy}$ is the Dyson sum, which contains terms of all orders but selects only those Feynman diagrams whose loops are arranged daisy-chain-like, shown in Eq.~\eqref{Fig-MTMovFeynman}:
\begin{subequations}
 \label{Fig-MTMovFeynman}
\begin{align}
\langle\psi_j(t)\widetilde\psi_0(0)\rangle_\text{Dy}\quad\corresponding&\quad \begin{tikzpicture}[>=Latex,line width=1.5pt]
  \draw[decorate, decoration={snake},color=blue] (-1.3,0) -- (1.3,0);
 \end{tikzpicture}\label{Eq-Dyson-01}\\
 &\quad\begin{tikzpicture}[>=Latex,line width=1.5pt,baseline=-0.2cm]
 \draw (0,0.3) node {$+$};
  \draw[color=red] (-0.5,0.1) .. controls (-0.5,-0.65) and (0.5,-0.65) .. (0.5,0.03);
 \draw[decorate, decoration={snake},color=blue] (-1.3,0) -- (1.3,0);
 \end{tikzpicture}\label{Eq-Dyson-02}\\
  &\quad\begin{tikzpicture}[>=Latex,line width=1.5pt,baseline=-0.1cm]
 \draw (0,0.3) node {$+$};
  \draw[color=red] (-0.7,-0.08) .. controls (-0.7,-0.6) and (0.0,-0.6) .. (0.0,-0.07);
  \draw[color=red] (0.0,-0.07) .. controls (0.0,-0.6) and (0.7,-0.6) .. (0.7,-0.05);
 \draw[decorate, decoration={snake},color=blue] (-1.3,0) -- (1.3,0);
  \draw (2,0) node {$+\quad\dots$};
 \end{tikzpicture}\label{Eq-Dyson-03}
\end{align}    
\end{subequations}  
\textit{A priori}, it is not clear which truncation of the path integral is a good approximation of the observable. However, a good agreement of the approximate result with experimental data indicates that the processes which were not included in the calculation rarely occur under experimental conditions.

The second and third approximation for the average filament growth speed are
\begin{subequations}\label{Eq-speed-one-loop}
\begin{align}
\langle v\rangle_{2}=&h(\lambda\zeta-\tau)\biggl(1-\underbrace{\lambda\Bigl(\frac{1}{4\pi DR}-\frac{h|\lambda\zeta-\tau|}{8\pi D^2}\Bigr)}_{\text{Loop correction term}}\biggr),\label{Eq-speed-one-loop-01}\\
\langle v\rangle_\text{Dy}&=\frac{h(\lambda\zeta-\tau)}{1+\lambda\Bigl(\frac{1}{4\pi DR}-\frac{h|\lambda\zeta-\tau|}{8\pi D^2}\Bigr)}.
 \end{align} 
\end{subequations}
The approximation $\langle v\rangle_0$ is given by $\langle v\rangle_2$ without the loop correction term.
The derivation of Eq.~\eqref{Eq-speed-one-loop} is outlined in the Supplementary Material, in Section~A5. 

The loop correction term takes into account that, due to diffusive transport, the reaction-limited speed is reduced further because the locally depleted particles have to reach the reaction sphere of radius $R$. Considering the first term of the loop correction, if the diffusion is strong, the chemical reactions are less hindered by slow transport and the growth speed is closer to its reaction limit. This diffusion correction is itself corrected in the second term of the loop correction which describes how quickly the tip reaches regions that are less depleted. It predicts a non-linear dependence of the growth speed $\langle v\rangle$ on the bulk particle density $\zeta$, which so far has not been observed experimentally. Therefore, we assume $\frac{1}{4\pi DR}\gg\frac{h|\lambda\zeta-\tau|}{8\pi D^2}$ and ignore the latter in the following. Thus, given the diffusion-limited growth coefficient $4\pi DR$ and the observed $k_\text{on}$ and $k_\text{off}$, Eq.~\eqref{Eq-Experiment-Speed}, we can calculate the reaction-limited $\lambda$ as well as the offset $\tau$\begin{align}\label{Eq-RD-rates}
\lambda=k_\text{on}\frac{4\pi DR}{4\pi DR-k_\text{on}}\quad\text{and}\quad\tau=k_\text{off}\frac{4\pi DR}{4\pi DR-k_\text{on}},
\end{align}
which is equivalent to the results in \cite{Noyes1961} for general reaction-diffusion systems.
It follows, that the observed growth coefficient $k_\text{on}$ is smaller than the reaction- and the diffusion-limited growth coefficients, i.e. $\frac{k_\text{on}}{\lambda}<1$ and $\frac{k_\text{on}}{4\pi DR}<1$, as shown in Fig.~\ref{Fig-Reaction-vs-diffusion-scheme}.

Furthermore, there is a bulk particle density $\zeta_\times$ at which diffusion becomes the defining limitation in comparison to reaction (Fig.~\ref{Fig-Reaction-vs-diffusion-scheme}, B):\begin{align}
\zeta_\times=\frac{k_\text{off}}{2k_\text{on}-4\pi DR}.
\end{align}   
For $k_\text{off}=0$, if $4\pi DR>2k_\text{on}$, then growth is reaction-limited otherwise it is  diffusion-limited.

One of the open questions for microtubule and actin filament growth is the origin of large fluctuations \cite{Gardner2011,Fujiwara2002}. In part, they can be explained as an overlap of the fluctuations of the reaction and diffusion processes.  This is quantified by the effective diffusion constant (derived in SI, Sec.~A6)\begin{align}\label{Eq-effect-Diff-2}
D_{\text{eff},2}=&h^2(\lambda\zeta+\tau)\left(1-\frac{\lambda}{4\pi DR}\right)+\lambda\frac{h|\lambda\zeta-\tau|}{2D\pi},
\end{align}
which is larger than the reaction-limited effective diffusion in Eq.~\eqref{Eq-R-Limit}.

In conclusion, the field theoretic model allows us to calculate how the reaction limit of microtubule and actin filament growth is undermined due to imperfect diffusive supply of tubulin or actin. It also predicts larger growth fluctuations compared to a model purely based on reactions.  Given the overlap of fluctuations due to chemical reactions and diffusive transport, it is likely for the filament growth speed to exhibit correlations in time, the study of which would be compelling for future research. 

The authors thank Robert Endres, Guillaume Salbreux and Thomas Surrey for very helpful discussions.
\newpage
\bibliographystyle{amsplain}
\bibliography{references.bib}
\end{document}


\maketitle
\section{Conventions}
\label{A-Introduction}
In the following sections, many details of the calculations whose results are shown in the main text are presented. The field-theoretic calculations use the following conventions for the Fourier transformations:\begin{align}\label{A-Eq-Fourier}\begin{split}
\varphi(\omega,k):=\mathcal{F}[\varphi(t,x)](\omega,k)=&\int\limits_{\mathbb{R}^4}\varphi(t,x)e^{i\omega t-ikx} \plaind^3x \plaind t,\\
\mathcal{F}^{-1}[\varphi(\omega,k)](t,x)=&\int\limits_{\mathbb{R}^4}\varphi(\omega,k)e^{-i\omega t+ikx} \dbar^3k \dbar\omega,\\
\psi(\omega,k):=\mathcal{F}[\psi_j(t)](\omega,k)=&\int\limits_{\mathbb{R}}\sum\limits_{j\in\mathbb{Z}^3}\psi_j(t)e^{i\omega t-ikhj}\plaind t,\\
\mathcal{F}^{-1}[\psi(\omega,k)](\omega,k)=&h^3\int\limits_{\mathbb{R}}\int\limits_{[0,\frac{2\pi}{h}]^3}\hspace{-0.3cm}\psi(\omega,k)e^{-i\omega t+ikhj}\dbar^3k\dbar\omega,
\end{split}\end{align}
with $\dbar\omega=\plaind\omega/(2\pi)$. The shorthand $\deltabar(\omega-\omega')=2\pi\delta(\omega-\omega')$, where $\delta(\cdot)$ is the Dirac $\delta$-function is also used. Furthermore, $\Theta(t)$ denotes the Heaviside function.

\section{Master equation}
\label{A-MasterEq}
In order to model filament growth as a field theory, we follow the approach by \textit{Doi} \cite{Doi1976} and \textit{Peliti} \cite{Peliti1985}. Our model for filament growth is described by a master equation. 
It includes six processes: \begin{itemize}
\item[(1)] diffusion of actin or tubulin particles (diffusion constant $D$), 
\item[(2)] tubulin or actin absorption by the filament  tip (incorporation constant $\lambda$) and subsequent movement of the tip in $z$ direction by one step on the lattice, 
\item[(3)] tubulin or actin release (rate $\tau$) and subsequent movement of the tip in $-z$ direction by one step on the lattice, 
\item[(4)] creation of tubulin or actin (constant $\gamma$), 
\item[(5)] extinction of tubulin or actin (rate $r$), 
\item[(6)] extinction of the filament tip (rate $\epsilon$). 
\end{itemize}

The master equation describes in continuous time and on a discrete spatial lattice $h\mathbb{Z}^3$ how many microtubule or actin filament tips ($m_j\in\mathbb{N}_0$) and how many tubulin or actin particles ($n_j\in\mathbb{N}_0$) are at position $j\in\mathbb{Z}^3$. Let $\{m\}$ denote the entire filament tip occupation configuration in $h\mathbb{Z}^3$, and $\{n\}$ denote the respective tubulin/actin population. Then, we denote by $\mathcal{P}(\{m\},\{n\},t)$ the probability for the system to be in this configurations at time $t$. Furthermore, we use the shorthand $1_j$ for occupation of one filament tip / particle at position $j$. The master equation is:\begin{align}\label{A-Eq-Master}\begin{split}
&\hspace{-1cm}\partial_t\mathcal{P}(\{m\},\{n\},t)=\sum\limits_{j\in\mathbb{Z}^3}\biggl[\\
(1)\hspace{0.5cm}&D\hspace{-0.2cm}\sum\limits_{|i-j|=1}\Bigl((n_j+1)\mathcal{P}(\{m\},\{n+1_j-1_i\},t)-n_j\mathcal{P}(\{m\},\{n\},t)\Bigr)\\
(2)\hspace{0.5cm}&+\lambda\Bigl((m_j+1)(n_j+1)\mathcal{P}(\{m+1_j-1_{j+e_z}\},\{n+1_j\},t)-m_jn_j\mathcal{P}(\{m\},\{n\},t)\Bigr)\\
(3)\hspace{0.5cm}&+\tau\Bigl((m_j+1)\mathcal{P}(\{m_j+1_j-1_{j-e_z}\},\{n-1_j\},t)-m_j\mathcal{P}(\{m\},\{n\},t)\Bigr)\\
(4)\hspace{0.5cm}&+\gamma\Bigl(\mathcal{P}(\{m\},\{n-1_j\},t)-\mathcal{P}(\{m\},\{n\},t)\Bigr)\\
(5)\hspace{0.5cm}&+r\Bigl((n_j+1)\mathcal{P}(\{m\},\{n+1_j\},t)-n_j\mathcal{P}(\{m\},\{n\},t)\Bigr)\\
(6)\hspace{0.5cm}&+\epsilon\Bigl((m_j+1)\mathcal{P}(\{m+1_j\},\{n\},t)-m_j\mathcal{P}(\{m\},\{n\},t)\Bigr)\biggr].\end{split}
\end{align}

\section{Second Quantized Model}
\label{A-SQM}
As outlined by \textit{Doi} \cite{Doi1976}, the classical many particle equation~\eqref{A-Eq-Master} can be transformed into a second quantized version. For this, time-independent occupation states $|\{m\},\{n\}\rangle$ are introduced, which represent the particle configuration, i.e. they tell us where how many filament tips ($\{m\}$) and tubulin or actin particles ($\{n\}$) are found. Furthermore, ladder operators $a_j$, $a_j^\dagger$ for filament tips and $b_j$, $b_j^\dagger$ for tubulin or actin are introduced.  Their commutation rules are $[a_j,a_i^\dagger]=[b_j,b_i^\dagger]=\delta_{ij}$. All other commutators are zero. Their action on occupation states is defined as\begin{align}
a_j|\{m\},\{n\}\rangle=&m_j|\{m-1_j\},\{n\}\rangle,& a_j^\dagger|\{m\},\{n\}\rangle=&|\{m+1_j\},\{n\}\rangle,\\
b_j|\{m\},\{n\}\rangle=&n_j|\{m\},\{n-1_j\}\rangle,& b_j^\dagger|\{m\},\{n\}\rangle=&|\{m\},\{n+1_j\}\rangle.
\label{A-Eq-Master-Equation}
\end{align}

The state of the system is defined as \begin{align}
|\phi(t)\rangle=\sum\limits_{\{m\},\{n\}}\mathcal{P}(\{m\},\{n\},t)|\{m\},\{n\}\rangle.
\end{align}

Using Eq.~\eqref{A-Eq-Master}, the time derivative of $|\phi(t)\rangle$ can be written as\begin{align}\begin{split}
\partial_t|\phi(t)\rangle=&\sum\limits_{j\in\mathbb{Z}^3}\biggl[D\hspace{-0.2cm}\sum\limits_{|i-j|=1}\Bigl(b_i^\dagger b_j-b_j^\dagger b_j\Bigr)\\
\hspace{0.5cm}&+\lambda\Bigl(a^\dagger_{j+e_z}a_jb_j-a_j^\dagger a_j b_j^\dagger b_j\Bigr)+\tau\Bigl(a^\dagger_{j-e_z}a_jb_j-a_j^\dagger a_j\Bigr)\\
\hspace{0.5cm}&+\gamma\Bigl(b_j^\dagger-1\Bigr)+r\Bigl(b_j-b_j^\dagger b_j\Bigr)+\epsilon\Bigl(a_j-a_j^\dagger a_j\Bigr)\biggr]|\phi(t)\rangle.\end{split}
\label{Second-quantized-eq}\end{align}

As described by \textit{Peliti} \cite{Peliti1985}, the second quantized state equation~\eqref{Second-quantized-eq} can be transformed into a field theory in path integral formulation. The result is presented in the main text of the article, Eqs.~(4),~(5),~(6), and~(7). 

\section{Propagators}
In order to calculate moments of observables, the action in Eqs.~(4),~(5),~(6), and~(7) is Fourier transformed, using the convention from Eq.~\eqref{A-Eq-Fourier}. In Fourier-space, the  propagators from Eqs.~(4),~(5),~(6) are represented by lines in Feynman diagrams:\begin{align}
\begin{tikzpicture}
\draw[color=red,line width=1.5pt] (-1,-0) -- (1.0,0);
\end{tikzpicture}\quad&\corresponding\quad \frac{\deltabar(\omega+\omega_0)\deltabar^3(k+k_0)}{-i\omega+Dk^2+r},\\
\begin{tikzpicture}
\draw[decorate, decoration={snake},color=blue,line width=1.5pt] (-1,-0) -- (1.0,0);
\end{tikzpicture}\quad&\corresponding\quad \frac{\deltabar(\omega+\omega_0)\deltabar_c^3(k+k_0)}{-i\omega+\lambda\zeta(1-e^{-ihk})+\tau(1-e^{ihk})+\epsilon},
\end{align}
where incoming frequencies and momenta carry the subscript $0$, while outgoing frequencies and moment do not have a subscript.

As the observables are probability distributions, after the Fourier transformation, the observables becomes their moment generating functions. Thus, moments and variances can be calculated as derivatives of the moment generating functions, evaluated at zero. This connection is used to calculated the expected filament growth speed and variance in the following sections.
\section{Expected microtubule growth length}
\label{A-Exp} 
In a first approximation, truncating the sum in Eq.~(8) at $\ell=0$, we calculate the expected filament tip position without any loop corrections:\begin{align}\begin{split}
\langle hj_z\rangle_0=&i\partial_{k_z}\int\limits_\mathbb{R}\langle\psi(\omega,k)\widetilde\psi(\omega',k')\rangle_0\dbar\omega\Bigr|_{k_z=0}\\
=&i\partial_{k_z}\int\limits_{\mathbb{R}}\frac{e^{-i\omega t}\dbar\omega}{-i\omega+\lambda\zeta(1-e^{-ih{k_z}})+\tau(1-e^{ih{k_z}})+\epsilon}\Bigr|_{k_z=0}\\
&=h(\lambda\zeta-\tau)te^{-\epsilon t}\Theta(t).\end{split}
\end{align}

If we now let the creation and extinction coefficients tend to zero while keeping their ratio $\zeta=\gamma/r$ constant, we find\begin{align}
\lim\limits_{\epsilon\rightarrow0}\langle hj_z\rangle_0=h(\lambda\zeta-\tau)t\Theta(t).
\end{align}
Hence, in this approximation, the filament growth speed is $\langle v\rangle_0=\langle hj\rangle_0/t=h(\lambda\zeta-\tau)$, which is the reaction-limited growth speed, shown in Eq.~(2) in the main article.

For the first correction $\langle v\rangle_2$, the process which is represented by the one-loop Feynman diagram, Eq.~(10b), has to be considered also:\begin{align}\begin{split}\label{A-Loop-MT}
\langle hj_z\rangle_2=&\langle hj_z\rangle_0+i\partial_{k_z}\int\limits_{\mathbb{R}^5}\langle\psi(\omega,k)\widetilde\psi(\omega',k')\rangle_1\dbar\omega\dbar\omega_b\dbar k_b\Bigr|_{k_z=0}\\
=&\langle hj_z\rangle_0+i\partial_{k_z}\hspace{-0.2cm}\int\limits_{\mathbb{R}^5}\frac{\lambda(e^{-ihk_z}-1)(\tau e^{ihk_z}-\lambda\zeta)e^{-i\omega t}}{(-i\omega+\lambda\zeta(1-e^{-ihk_z})+\tau(1-e^{ihk_z})+\epsilon)^2}\\
&\frac{\dbar^3 k_b\dbar\omega\dbar\omega_b}{(i(\omega_b-\omega)+\lambda\zeta(1-e^{-ih(k-k_b)_z})+\tau(1-e^{ih(k-k_b)_z})+\epsilon)(-i\omega_b+Dk_b^2+r)}\Bigr|_{k_z=0}\\
\approx&\langle hj_z\rangle_0+-h(\lambda\zeta-\tau)t\underbrace{\lambda\Bigl(\frac{\Lambda}{2\pi^2 D}-\frac{h|\lambda\zeta-\tau|}{8\pi D^2}\Bigr)}_{\text{Loop correction term}}\Theta(t),
\end{split}\end{align}
where $\omega_b$ and $k_b$ are the loop's free frequency and momentum:\begin{align}
\text{Loop}(\omega,k)=\int\limits_{\mathbb{R}^4}\frac{\dbar^3 k_b\dbar\omega_b}{(i(\omega_b-\omega)+\lambda\zeta(1-e^{-ih(k-k_b)_z})+\tau(1-e^{ih(k-k_b)_z})+\epsilon)(-i\omega_b+Dk_b^2+r)}
\end{align}
Here, we approximated $(1-e^{\pm ih(k_b-k)_z})$ by $\mp ih(k_b-k)_z$, as well as $(1-e^{\pm ihk_z})$ by $\mp ihk_z$. Then, a new variable $\tilde k_b$ is introduced, for which the $z$-component is shifted by $ih(\lambda\zeta-\tau)/(2D)=k_{b_z} -\tilde k_{b_z}$. Furthermore, we let $r,\epsilon$ tend to zero and we introduced a cutoff for the $\tilde k_b$ integral such that $|\tilde k_b|<\Lambda$. This cutoff $\Lambda$ is identified as $\Lambda=\pi/(2R)$, where $R$ is the radius of the reaction sphere.

Then, the calculation splits into a steady state part (shown here) and a relaxation part, where the latter tends to zero for large times $t$. Here, we focus on steady state solutions and omit the relaxation part. The average expected first correction to the growth speed is $\langle v\rangle_2=\langle hj_z\rangle_2/t$, see Eq.~(11) of the main article. Both corrections are graphically represented by Feynman diagrams in Eq.~(10a) and~(10b) of the main article.

The Dyson sum is a geometric sum over loop corrections where all the loops are daisy-chain-like, see Eq.~(10) for the first three diagrams in the series. The associated expected position is\begin{align}\begin{split}
\langle hj_z\rangle_\text{Dy}=&\langle hj_z\rangle_0+i\partial_{k_z}\hspace{-0.2cm}\int\limits_{\mathbb{R}^5}\frac{\lambda(e^{-ihk_z}-1)(\tau e^{ihk_z}-\lambda\zeta)e^{-i\omega t}}{(-i\omega+\lambda\zeta(1-e^{-ihk_z})+\tau(1-e^{ihk_z})+\epsilon)^2}\cdot\frac{\text{Loop}(\omega,k)}{1-\lambda\text{Loop}(\omega,k)}\dbar\omega\Biggr|_{k_z=0}\\
\approx&\frac{h(\lambda\zeta-\tau)t}{1+\lambda\left(\frac{\Lambda}{2\pi^2 D}-\frac{h|\lambda\zeta-\tau|}{8\pi D^2}\right)}
\end{split}\end{align}
where $\text{Loop}(\omega,k)$ is the loop integral, i.e. the $\omega_b$ and $\tilde k_b$ integral over the third line of Eq.~\eqref{A-Loop-MT}. The speed is identified as $\langle v\rangle_\text{Dy}=\langle hj_z\rangle_\text{Dy}/t$.
\section{Variance of microtubule growth length}
\label{A-Var}
First, we calculate the mean square displacement of the filament growth in the approximation without loops, i.e. the sum in Eq.~(8) is truncated at $\ell=0$:
\begin{align}\begin{split}
\langle(hj_z)^2\rangle_0=&-\partial_{k_z}^2 \int\limits_\mathbb{R}\langle\psi(\omega,k)\widetilde\psi(\omega',k')\rangle_0\dbar\omega\Bigr|_{k_z=0}\\
=&h^2\Theta(t)e^{-\epsilon t}\Bigl((\lambda\zeta+\tau)t+(\lambda\zeta-\tau)^2t^2\Bigr).
\end{split}\end{align} 
The variance is equal to\begin{align}\begin{split}
\text{Var}_0(hj_z)=&\langle(hj_z)^2\rangle_0-\langle hj_z\rangle_0^2\\
=&h^2\Theta(t)(\lambda\zeta+\tau)t,
\end{split}\end{align}
where we took the limit $\epsilon\rightarrow0$. Hence, the variance of the average speed $\langle v\rangle_0$ decreases as $1/t$.\begin{align}
\text{Var}_0(v)=\frac{\text{Var}_0(hj_z)}{t^2}=\frac{h^2(\lambda\zeta+\tau)}{t}.
\end{align}
These results represent the reaction-limited behaviour. As the variance of the growth length is linear in time, an effective diffusion constant $D_{0,\text{eff}}=h^2(\lambda\zeta+\tau)$ can be associated. This observable is often considered as a sign for the fluctuations of the growth process. However, it characterises the reaction-limited fluctuations. In order to estimate the influence of additional, diffusive fluctuations, at least the first loop correction has to be calculated:
\begin{align}
\langle (hj_z)^2\rangle_2=&\langle(hj_z)^2\rangle_0-\partial^2_{k_z}\hspace{-0.2cm}\int\limits_{\mathbb{R}}\frac{\lambda(e^{-ihk_z}-1)(\tau e^{ihk_z}-\lambda\zeta)e^{-i\omega t}}{(-i\omega+\lambda\zeta(1-e^{-ihk_z})+\tau(1-e^{ihk_z})+\epsilon)^2}\text{Loop}(\omega,k)\dbar\omega\Bigr|_{k_z=0}\\
\approx&\langle(hj_z)^2\rangle_0-\Theta(t)\left(h^2(\lambda\zeta+\tau)t+2h^2(\lambda\zeta-\tau)^2t^2\right)\underbrace{\lambda\left(\frac{\Lambda}{2\pi^2 D}-\frac{h|\lambda\zeta-\tau|}{8\pi D^2}\right)}_{\text{Loop correction term}}+\underbrace{\lambda\frac{h|\lambda\zeta-\tau|}{2D\pi}t}_{\text{diffusive term}},
\end{align}
where the same approximation as for Eq.~\eqref{A-Loop-MT} where used, and the extinction rates $\epsilon$ and $r$ were set to zero. The extra 'diffusive term' describes the additional fluctuation that go beyond the purely chemical reactions fluctuations, which are described by a Skellam distribution \cite{Skellam1946}. 

In order to calculate the variance, the expected growth length squared has to be subtracted. However, as we are only considering the first correction, we omit all terms in which the 'Loop correction term' appears in a higher than linear order and find the result shown in Eq.~(14) of the main article.

\bibliographystyle{amsplain}
\bibliography{references.bib}